\documentstyle[11pt,newpasp,twoside,epsf]{article} 
\def\edcomment#1{\iffalse\marginpar{\raggedright\sl#1\/}\else\relax\fi} 
\reversemarginpar 

\begin{document} 
\title{Accurate Internal Proper Motions of Globular Clusters}

\author{Luigi R.\ Bedin and Giampaolo Piotto} 
\affil{Dipartimento di Astronomia, Universit\`a di Padova, 
Vicolo dell'Osservatorio, 2, I-35122, Padova, Italy}
\author{Jay Anderson} 
\affil{Rice University, USA} 
\author{Ivan R.\ King} 
\affil{Astronomy Department, University\ of Washington, Box 351580, Seattle,
WA 98195-1580, USA} 

\begin{abstract} 
We have  undertaken a  long  term  program to measure  high  precision
proper motions  of nearby Galactic  globular cluster  (GC) stars using
multi-epoch  observations with the  WFPC2 and the ACS cameras on-board
the Hubble Space  Telescope. The proper motions are  used to study the
internal cluster kinematics, and to obtain accurate cluster distances.
In this paper, we also show how the proper motions  of the field stars
projected in the direction of the studied clusters can  be used to set
constraints on the Galaxy kinematics.
\end{abstract}

\section{Proper Motions in the Field of M4}

Some of us have recently developed  a method (Anderson and King, 2000,
2003), based on multi-epoch HST images, which allows us to derive very
accurate proper motions.   The key  to our success  is the  removal of
systematic errors  of many types from   our astrometry.  Presently, we
are able  to measure the  position of a  well exposed star in a single
image with a precision of 0.02 pixel, i.e.\ 2 mas in the WFs, 1 mas in
the PC.  Multiple observations  allow to reduce  this uncertainty of a
factor $\sqrt{N}$, where N is the number of frames.

\subsection{The Internal Proper Motion of M4 and its Distance}

The  internal proper motions of  M4 is  $\sim$0.50 mas/yr.  This means
that with our 8 frames in each of two epochs, and with a time baseline
of $\sim$5 yrs, we can reach a precision  of $\sim$0.2 mas/yr.  We can
then subtract, in  quadrature, the distribution of  the  errors to the
observed  distribution, and  have a precise   measure of  the internal
dipersion   of proper motions.   Figure\    1 shows some   preliminary
results. We plan  to furtherly improve the  proper motions of M4 stars
by the reducing the third  epoch WF@ACS images  which overlap with our
fields.

We   have  recently undertaken a    program aimed at  obtaining radial
velocities for stars located  in the same  radial interval and  in the
same magnitude  range of  the  proper  motion data in  M4  (and  other
clusters).    A comparison of the proper   motion  dispersion with the
radial velocity dispersion will lead  to the direct measurement of the
GC distance.  The major source of uncertainty being the sampling error
(in first  approximation $1/\sqrt{2N_{tot}}$,  where $N_{tot}$ is  the
total number of measured stars), which gives an error 1.3 $\%$ for the
3,000 stars  we plan to observe in  each cluster  with the multi-fiber
high resolution  facility FLAMES@VLT.  The statistical limitation will
be set by the number of radial  velocities, as we have accurate proper
motions for a large number of stars.

%%%%%
\begin{figure}
\vspace{-.1in}
\plotone{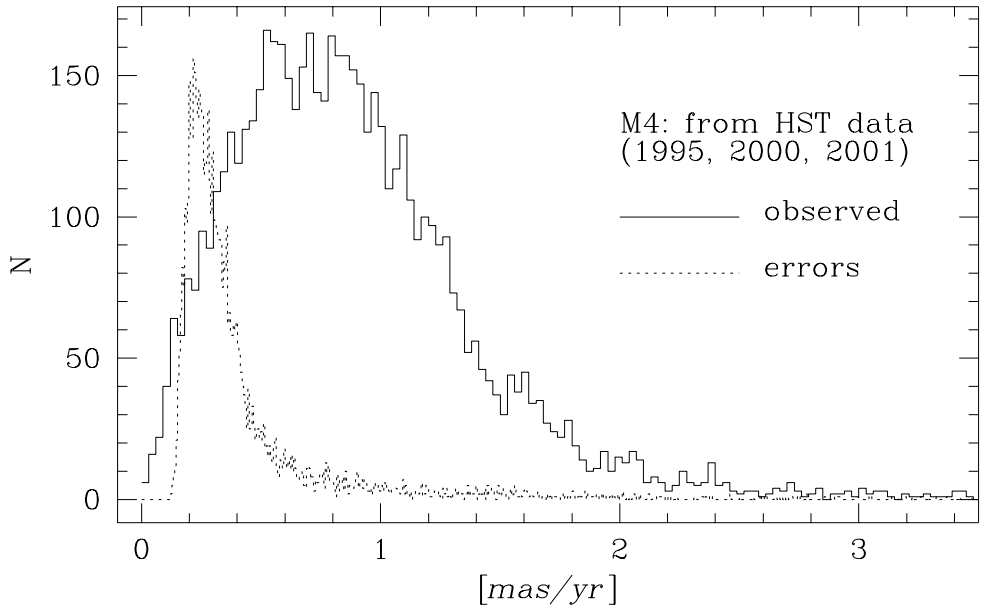}
\caption{The preliminary internal proper motion in M4 (solid line)
is compared with the  astrometric errors (dotted line).  }
\end{figure} 
%%%%%

\subsection{Measure of the Galactic Constant $\Theta_0/R_0$}

An  obvious  by-product of  our   proper   motion measurements  is   a
decontamination  of the cluster sample  from the  field stars (King et
al.\ 1998, Bedin et  al.\ 2001).  There  is a number of  studies which
can be undertaken with the  motions of the fields  objects that we get
for  free. We  give  a   brief description   of  an example  of   such
applications.

%%%%%%%%%%%%%%%%%%%%%%%%%%%%%%%%%%%%%%%%%%%%%%%%%%%%%%%%%%%%%%%%%%%%%%

%
In Bedin et al.\ (2001),  two $HST$ deep  observations of the globular
cluster M4, separated by a time base line of  $\sim$ 5 yrs, allowed us
to obtain  a pure  sample of the   cluster main  sequence stars.   The
identification of an extra-Galactic point source enables us to use the
proper motions of field stars (which were junk in  Bedin et al.\ 2001)
to measure a fundamental parameter of the Galaxy.
M4 is a globular  cluster projected on  the edge of the Galactic bulge
($\ell\simeq-9^\circ$, $b\simeq16^\circ$).   We   expect only  a small
number of foreground disk  stars in our  fields, but in the background
we look  through the edge  of the  bulge at a   height of $\sim2$ kpc.
Although at such heights  the density of  the bulge is rather low, the
volume we are  probing is sizable, so that   we see a  large number of
bulge members.
%
%%%%
\begin{figure}[ht!]
\plotone{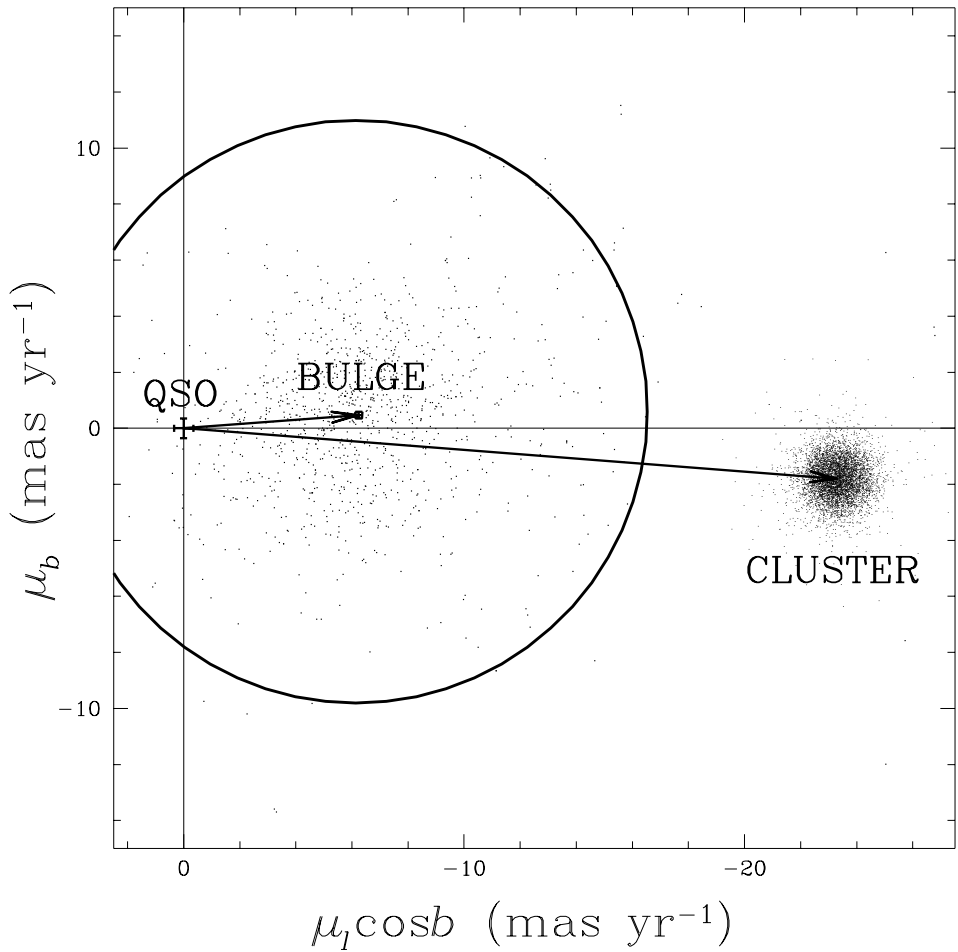}
\caption{Vector-point diagram of all the independent  measurements  of 
the in Galactic proper  motions.  The arrows  indicate the mean motion
of the cluster and the bulge  with respect to an extragalactic source.
}
\label{fig:bulgeAB} 
\end{figure} 
%%%%
Their absolute proper motion (pm) is just  the reflection of the Sun's
angular velocity  with respect to  that  point; from  that pm  we  can
derive the value of the angular velocity of the local standard of rest
(LSR)  with respect to the  Galactic center,  which is the fundamental
Galactic-rotation   constant  $A-B=\Theta_0/R_0$     (cf.\    Kerr  \&
Lynden-Bell 1986).

To  derive this value  we need to:\ (1)  find the mean distance of the
bulge stars whose motion we are observing, (2) correct the observed pm
for  the velocity of the Sun  with respect to the  LSR, and (3) relate
the corrected pm to the  angular velocity of  the LSR with respect  to
the Galactic center.

For the distance of the  bulge stars that  we are observing, we assume
the following working hypotheses:
(1) Disk and halo stars are a negligible component of the field stars
    in our M4 images, i.e., the field stars are mainly bulge members.
(2) The bulge stars on our line of sight are part of a spherical
    spatial  distribution around the  Galactic center. 
(3) Our observations go deep enough that we do not loose stars on the
    far side of the bulge.
From these assumptions, it follows that we can express the distance of
the centroid of the bulge stars along our line of sight (we will refer
to  it as the bulge) as  a geometrical constant$\times$the distance of
the Sun from the Galactic center.
This distance is
$R=R_0~\cos\ell \cos b.$
If we take $R_0=8.0$ kpc, then $R=7.6$ kpc.\\
The absolute  motion of the bulge  --corrected for the  Sun's peculiar
motion-- allows us   to get a  direct  estimate of  the  Oort-constant
difference $(A-B)$, which is related to the transverse velocity of the
LSR ($\Theta_0$) and its Galactocentric distance ($R_0$), according to
equations presented in Bedin et al.\ (2003a; in Bedin et al.\ 2003b we
will give   more details and a  more   concise formulation  leading to
identical results),     we  get:   $(A-B)  \pm   \sigma_{(A-B)}$   $=$
$\Theta_0/R_0 \pm  \sigma_{\Theta_0/R_0}$  $=$  $27.6 \pm  1.7$  ${\rm
km/s~{\rm kpc}}$.  The quoted error is  internal and corresponds to an
uncertainty of 7$\%$.
\subsection{Galactic Model }
Finally, we mention that the field  star data can  be used to probe in
depth the Galactic  structure  and stellar content.  We  are presently
collaborating with the  research group at the  University of Pisa on a
project aimed  at the definition of  a Galactic  model.  An example on
how the field  star data  from proper  motions   as those  in King  et
al. (1998) and Bedin   et al. (2001)  can   be used  in this  kind  of
investigation is in Castellani et al. (2001).
\end{document}